\documentclass[useAMS,usenatbib]{mn2e}
\usepackage{graphicx}
\usepackage{latexsym,epsfig}
\usepackage{longtable,rotating}

\newcommand{\apj}{\mbox{ApJ}}
\newcommand{\aap}{\mbox{A\&A}}
\newcommand{\mnras}{\mbox{MNRAS}}

\newcommand{\aj}{\mbox{AJ}}
\newcommand{\araa}{\mbox{ARA\&A}}
\newcommand{\pasa}{mbox{Publ. Astron. Soc. Australia}}
\newcommand{\pasp}{mbox{PASP}}

\usepackage{lscape}
\usepackage{array,arydshln}

\title[Hyades and Sirius streams]{Chemical compositions of giants in the Hyades
And Sirius superclusters }
\author[Ramya, Reddy and Lambert]{P. Ramya,$^{1,2}$\thanks{E-mail:
ramyap09@gmail.com} Bacham E. Reddy$^{1}$ and David L. Lambert$^{2}$ \\
$^{1}$Indian Institute of Astrophysics, Bengaluru, India-560034\\
$^{2}$The W.J. McDonald Observatory \& Department of Astronomy,University of Texas at Austin, Austin, TX 78712, USA  }

\begin{document}

\date{Accepted ..... Received .... ; in original form ....}

\pagerange{\pageref{firstpage}--\pageref{lastpage}} \pubyear{...}

\maketitle

\label{firstpage}

\begin{abstract}
An abundance analysis for 20 elements from Na to Eu is reported  for 34 K
giants from the Hyades supercluster and for 22 K giants from the Sirius 
supercluster. Observed giants were identified as highly probable members 
of their respective superclusters by Famaey et al. (2005, A\&A, 430, 165).
Three giants each from the Hyades and Praesepe open clusters were 
similarly observed and analysed. Each supercluster shows a range in
metallicity: $-0.20 \leq$ [Fe/H] $\leq +0.25$ for the Hyades supercluster
and $-0.22 \leq $ [Fe/H] $\leq +0.15$ for the Sirius supercluster with the
metal-rich tail of the metallicity distribution of the Hyades supercluster 
extending beyond that of the Sirius supercluster and spanning
the metallicity of the Hyades and Praesepe cluster giants. Relative
elemental abundances [El/Fe] across the supercluster giants are
representative of the Galactic thin disc as determined from giants in
open clusters analysed in a similar way to our approach. Judged by
metallicity and age, very few and likely none of the giants in these
superclusters originated  in an open cluster:  the pairings
include the Hyades supercluster with the Hyades - Praesepe open
clusters and the Sirius supercluster with the U Ma open cluster. 
Literature on main sequence stars attributed to
the two superclusters and the possible relation to the
associated open cluster is reviewed. It is suggested that
the Hyades supercluster's main sequence population contains few
stars from the two associated open clusters. As suggested by some
previous investigations, the Sirius supercluster,
when tightly defined kinematically, appears to be well populated by 
stars shed by the U Ma open cluster.
\end{abstract}

\begin{keywords}
 stars: abundances --- stars: moving groups--- Galaxy: kinematics and dynamics---Galaxy: disc
\end{keywords}

\section{Introduction}

Stellar superclusters are  overdensities in the  $(U,V,W)$ space defined by Galactic motions  where $U$, $V$ and $W$ are 
the radial, tangential and vertical stellar velocity component. Such an overdensity may be referred to also as a 
stellar moving group or  a stellar stream.  Modern  identifications of kinematic over-densities in the Galactic disc began with
Eggen (1958a,b,c) who suggested members of a moving group came from a dissolving open cluster (see Eggen 1996, and references therein). 
With evidence accumulating that stars within an  open cluster share a common chemical composition and age but stars within a 
stellar supercluster have a spread in composition and age, the perfect association of every supercluster with a dissolved open cluster 
is unlikely. Partial contamination from an open cluster remains a possibility. Origins of a  supercluster now include the idea that 
they may be composed of field stars given a common motion through dynamical perturbations arising from the Galaxy's spiral arms 
and/or the central bar (Antoja et al. 2010; Sellwood 2014). Finally, a supercluster may result, as do many stellar streams in 
the Galactic halo, from the accretion and disruption of a satellite galaxy but this would seem to be a remote possibility because 
the superclusters generally share the rotation of the Galactic disc.

In our first paper on superclusters (Ramya et al. 2016), we presented and discussed chemical compositions 
of giants identified with the Hercules stream by Famaey et al. (2005). Here, we extend our work to giants of 
two other superclusters - the Hyades and Sirius superclusters. The present pair differ from the Hercules stream 
in that their $(U,V,W)$ velocities  appear at first sight to be associated with  open clusters. The Hyades supercluster's 
possible association is with the Hyades and Praesepe open clusters. The Sirius supercluster is linked with the 
U Ma open cluster. In contrast, the Hercules stream has not been similarly tied to an open cluster. This difference 
between the present pair of superclusters and the Hercules stream may be reflected in the chemical compositions of 
stars in the three systems: some Hyades and Sirius supercluster stars may have chemical compositions  representative 
of the associated open cluster but  Hercules stream members are anticipated by theory  and shown by observation to 
have a spread in chemical compositions.

With the publication of trigonometrical parallaxes and proper motions from the {\it Hipparcos} satellite, the presence of  
stellar moving groups among local FGK dwarfs was rediscussed by several authors. Notably, Montes et al. (2001)  reconsidered 
five young kinematic groups with ages spanning from just 20 Myr up to about 600 Myr. The two oldest of the five kinematic groups 
are the focus of this paper: the Hyades supercluster at 600 Myr and the Sirius supercluster at 300 Myr, also known as the 
Ursa Major group.  The mean $(U,V,W)$ of the Hyades supercluster is similar to the mean motions of the Hyades and 
Praesepe open clusters. The mean $(U,V,W)$ of the Sirius supercluster is similar to the motion of the Ursa Major open cluster. 
(Montes et al. also identified three younger kinematic groups which are not considered here: the Castor moving group at 200 Myr,
the IC 2391 supercluster at 35-55 Myr and the Local Association at 20-150 Myr associated with several open clusters. 
The Local Association has also been called the Pleiades moving group). Further advances and discussion of moving groups are to 
be anticipated with the astrometry from the $\it Gaia$ satellite - see, for example, Kushniruk, Schirmer \& Bensby (2017). 
Wielen (1971) from the age distribution of open clusters within 1000 pc showed that 50\%  evaporated within 200 Myr, 10\% have 
a lifetime longer than 500 Myr and only 2\% survive for longer than 1000 Myr. Wielen's estimates suggest that significant 
evaporation may have occurred from the Hyades and U Ma open clusters: is this population now in their respective superclusters?

Our reexamination of the origins of the Hyades and Sirius superclusters is based on GK giants. An open cluster's contributions of 
dwarfs  and giants to the population of a supercluster may differ substantially.  For an  open cluster the  giants come from stars 
more massive than the cluster's present FGK dwarfs and the lifetime as a GK giant is much shorter than the lifetime of its 
main sequence progenitor and  much shorter than the lifetime of the remaining FGK dwarfs. In addition,  the initial mass function 
increases with decreasing mass and this further reduces the number of GK giants relative to the number of FGK dwarfs, as is evident 
from the scarcity of giants in open clusters. The  corollary is that a search among a supercluster for evaporated stars from a 
parent open cluster may be more productive  using main sequence stars than giant stars. This conclusion is subject to qualification 
on account of differing degrees of dilution of the two samples by `field' giant and `field' main sequence stars, respectively.

In this paper, we determine the chemical compositions for giant stars identified by Famaey et al. (2005) as belonging to the Hyades 
and the Sirius superclusters,  compare their compositions with previously reported results for these clusters and superclusters and 
attempt to assess the contributions of the associated open clusters to the superclusters. 
 
\section{The samples}

Famaey et al. (2005) applied a maximum-likelihood method to  kinematic data assembled from the {\it Hipparcos} catalogue in
order to identify structures in the $(U,V)$ plane. Six structures were isolated. About 80\% of the giants were placed within 
three structures: young giants with small velocity dispersions accounting for 10\% of the total sample, high-velocity giants 
from the thick disc and the halo accounting for a further 11\% and a smooth background in the $(U,V)$ space accounting for 
60\% of the {\it Hipparcos} sample. The remaining 20\% of the sample was divided between three groups with roughly equal 
populations but  distinctive motions: the Hyades-Pleiades supercluster with mean velocities $(U,V,W) = (-30.3, -20.3, -4.8)$, 
the Sirius  moving group with  mean velocities $(U,V,W) = (+6.5, +4.0, -5.8)$ and the Hercules stream with 
$(U,V,W) = (-42.1, -51.6, -8.1)$. (All quoted velocities are heliocentric values in km s$^{-1}$.)  Our present sample is drawn 
from those giants having a high probability of belonging to either the Hyades-Pleiades (here, the Hyades)  supercluster or the 
Sirius supercluster (aka, the U Ma group). Our sample of giants from the Hercules stream was discussed previously (Ramya et al. 2016).

In order to relate the chosen giants to the young stellar kinematic groups isolated by Montes et al., Figure 1 shows the 
velocities of the individual giants selected by us  from Famaey et al., the mean velocities of the Hyades supercluster and 
the Sirius supercluster from Famaey et al. and the mean velocities for the same kinematic groups from Montes et al.  The most 
recent wavelet analysis of supercluster locations in the $(U,V)$ plane from {\it Gaia DR1} parallaxes, {\it Tycho} proper motions 
and {\it RAVE} radial velocities (Kushniruk, Schirmer \& Bensby 2017) places the Hyades supercluster at $(-44,-18)$ with an 
elongation of about 10 km s$^{-1}$ in $U$. Kushniruk et al. note that previously published  $(U,V)$ centroids for the supercluster 
range from their own to $(-30,-15)$ by Antoja et al. (2012) and Bobylev \& Bajkova (2016). The mean velocity is approximately 
between the values given by Famaey et al. (2005) and Montes et al.  (2001)  in Figure 1. The Sirius supercluster is spread out 
in $U$ and $V$ with the leading component at $(0,8)$ according to Kushniruk et al. but the consensus from earlier estimates 
compiled by these authors place the $(U,V)$ velocities close to the location given by Famaey et al. (2005).

The two panels of Figure 1 show the clear separation of the Sirius supercluster (the Ursa Major group in Montes et al.'s  designation) 
from the Hyades supercluster members. A comparable separation is found  between the Sirius supercluster and the other kinematic groups
identified by Montes et al.: the Local Association, the IC 2391 supercluster and the Castor moving group.  The mean motions of the
Local Association associated with the Pleiades (and other open clusters),the IC 2391 and the Castor group  are shown in Figure 1.  
There is a difference of a few km s$^{-1}$ between the mean velocity of the Hyades and of the Sirius supercluster as found by 
Famaey et al. and by Montes et al. 

Montes et al. link their Hyades supercluster to the Hyades and Praesepe clusters. It is their Local Association which they 
tie to the Pleiades (and other) clusters and for which they offer the alternative name the Pleiades moving group. Famaey et al.
adopt the label Hyades-Pleiades supercluster implying a sampling of the Hyades and the Pleiades (and not the Praesepe)
clusters.  Not surprisingly, Figure 1 hints that the Famaey et al.'s mean velocities for the Hyades-Pleiades supercluster 
fall between Montes et al.'s values for the Hyades supercluster and the Local Association. It would appear from the distribution of 
our giants in Figure 1 that our Hyades sample is little contaminated by the Local Association (i.e., the Pleiades and other clusters) 
and the Castor moving group. Confusion with the IC 2391 group may arise when considering compositions.

Famaey et al. (2005) give membership probabilities for a giant to belong to the
Hyades and Sirius superclusters. Stars were selected by us if the probability of membership
was greater than 70 per cent with additional constraints 
including the colour cut off $V-I < 1.2$ to avoid M stars and that a star 
be observable from the W.J. McDonald Observatory during the observing run.
The final samples of giants with determined chemical composition include
34  from the Hyades supercluster  and 22 from the Sirius supercluster. Their
locations in the $(U,V)$ and $(W,V)$ planes are shown in Figure 1
with blue squares for the Hyades and green triangles for the Sirius supercluster
stars. Galactic velocities (U,V,W) computed from Gaia astrometry provide distributions 
within essentially identical to those in Figure 1 which are provided from HIPPARCOS astrometry, 
that is stream membership is not materially affected by our (ie., Famaey et al.'s) choice of HIPPARCOS astrometry.

In addition to members of the superclusters, we observed  giants $\gamma$, $\delta$ and $\epsilon$ Tau 
from the Hyades open cluster and giants HD 73598, 73665 and 73710 from the Praesepe open cluster. These 
stars are recognized in Figure 1 by orange circles for Hyades and orange diamonds for Praesepe giants.  
There are no giants recognized as belonging to the U Ma cluster.

\section{Observations and Abundance Analysis}

Stars in Table 1 for the Hyades supercluster and in Table 2 for the Sirius supercluster were observed with the 
Robert G. Tull coud\'{e} spectrograph  at the Harlan J. Smith 2.7 meter telescope at the W.J. McDonald
Observatory (Tull et al. 1995). Spectra cover the wavelength range 3800-10000 \AA\ but
longward of 5800 \AA\ coverage is incomplete because the free spectral
range of the echelle  exceeds the width of the CCD.  The spectral resolving power is
about 60000 and the signal-to-noise ratio of a typical spectrum is 100
or more over much of the spectrum. Wavelength calibration was provided by 
observation of a Th-Ar hollow cathode lamp. All reductions were carried
out with the software package {\it Image Reduction 
and Analysis Facility (IRAF)}.\footnote{ {\it IRAF} is distributed by the
National Optical Astronomy Observatory, which is operated by the Association
of Universities for Research in Astronomy (AURA) under cooperative
agreement with the National Science Foundation.}


Conversion of spectra to elemental abundances followed the
procedures described in our paper on the Hercules stream (Ramya et al.
2016). Model stellar atmospheres are taken from the Kurucz (1998) 
grid.\footnote{http://kurucz.harvard.edu/grids.html} The line list is an expanded  version of that given by
Ramya et al. (2016). The 2016 list  was extended by adding Fe\,{\sc I} and Fe\,{\sc ii} lines and now includes 
heavy elements  based on the line selection adopted by Reddy, Giridhar \& Lambert (2012, 2013). 
The equivalent width of these lines are measured manually from the spectrum and the abundances are calculated from individual lines 
using the 2014 version of the LTE spectral line analysis code MOOG (Sneden 1973) force-fitting the abundances to 
match the equivalent widths. Hyperfine corrections are applied for the elements Sc, V, Mn, Co, Ba and Eu using the 
{\it blends} driver of MOOG.
Table 3 lists the selected lines including their solar equivalent widths and the corresponding solar abundance. 
This Table  also gives the mean solar elemental abundance,  the mean solar abundance recommended by Asplund et al. (2009) 
and the (small) difference between the latter recommendation and the mean abundance from our selected lines. 
The five additional elements Y to Eu were added as tracers of neutron capture synthesis which are not included among the 
lighter elements considered in our previous paper on the Hercules stream. Addition of the five was in large part stimulated 
by our paper 'Prospecting for chemical tags among open clusters' (Lambert \& Reddy 2016).
The Y\,{\sc ii} 5289.81 \AA\ and Nd\,{\sc ii} 4989.92 \AA\ lines give systematically lower abundances at lower temperatures 
and, hence, are not considered in determining mean abundances for giants with effective temperatures cooler than 4700 K.

Model atmosphere parameters -- effective temperature $T_{\rm eff}$,
surface gravity $\log g$, microturbulence $\xi_t$ and metallicity
[M/H] -- are estimated in the customary fashion from the Fe\,{\sc i}
and Fe\,{\sc ii} lines. Results are provided in Tables 1 and 2 for the two
superclusters. Uncertainties are estimated to be $\Delta T_{\rm eff} =
\pm50$ K, $\Delta \log g = \pm 0.2$ dex, $\Delta \xi_t = \pm 0.2$ km s$^{-1}$
and $\Delta$ [M/H] = $\pm 0.1$ dex (see Ramya et al. 2016).

Photometry, as described by Ramya et al. (2016), provides a  check
on the spectroscopic parameters. Effective temperatures are obtained
from the  $(V-K)$ colour with uncertainties estimated to be
 $\pm 40$ K.
 The mean difference between
the spectroscopic and photometric temperature is $-49\pm141$ K and $+56\pm48$ K for the
Hyades and Sirius superclusters, respectively.  
Surface gravities were obtained through the web interface
for the PARAM code\footnote{http://stev.oapd.inaf.it/cgi-bin/param.1.3}
using the PARSEC isochrones (Bressan et al. 2012), as described by Ramya et al.
(2016). Surface gravities from the PARAM code differ slightly from
spectroscopic values: the mean difference is $0.02\pm0.09$ dex and $+0.04\pm0.08$ dex for
the Hyades and Sirius superclusters, respectively.

With the spectroscopic atmospheric parameters, elemental abundances were
estimated for all entries in the line list. Differential abundances
[El/H] are estimated using the solar abundances  from Asplund et al. (2009) 
and Table 3. Tables 4, 5 and 6 for the giants from the Hyades supercluster
give [Fe\,{\sc i}/H], [Fe\,{\sc ii}/H] and then
[El/H] with the line-to-line spread $\sigma$ and the number of lines used
for the abundance determination. Tables 7, 8 and 9 give the same information for
the giants of the Sirius supercluster. Ionization equilibrium (i.e.,
[El\,{\sc i}/H] $\equiv$ [El\,{\sc ii}/H]) is necessarily satisfied for
Fe but it may be noted that it is also closely satisfied for Ti where the
mean [Ti\,{\sc i}/H] $-$ [Ti\,{\sc ii}/H] is  $+0.04\pm0.05$ dex for both
superclusters.

Table 10 gives the atmospheric parameters and the
abundance information for the giants belonging to the Hyades and     
Praesepe clusters. Our discussion of the Hyades  supercluster relies partly  
on comparison of the compositions of the giants from the Hyades and Praesepe 
open clusters with those from the supercluster. Systematic abundance errors are 
expected to cancel almost exactly in this comparison. Inspection of Table 10 
shows that the spread in [Fe/H] among the trio from each cluster is 0.03 dex 
which should indicate the range of measurement uncertainty for giants in each 
supercluster. It is also of interest to consider how results in Table 10 compare 
with recent results in the literature for the same giants.

Literature on the Hyades cluster's metallicity is vast. We refer to a sample of recent studies. 
Dutra-Ferreira  et al. (2016) investigated the iron abundance of the three  giants and a sample 
of main sequence stars from the Hyades cluster. The authors used classical model atmospheres with 
two different ways of selecting the atmospheric parameters and two selections of Fe\,{\sc I} and Fe\,{\sc ii} lines. 
Our mean Fe abundance for the three giants is [Fe/H] = $+0.14$ or $\log$(Fe) = 7.58 since our line selection provides 
a  solar Fe abundance of 7.44. This Fe abundance  lies between the estimates by Dutra-Ferreira et al. for their 
two line lists and provided by a method similar to ours. Tabernero, Montes \& Gonz\'{a}lez Hern\'{a}ndez (2012)  
analysed Hyades open cluster and supercluster members including the giant $\epsilon$ Tau for which they obtained 
the Fe abundance 7.67, a value only 0.09 dex greater than our value.  Carrera \& Pancino (2011) analysed the three 
Hyades giants to obtain a mean Fe abundance of 7.61, also in good agreement with our result. There is a similar level 
of consistency for the Praesepe giants. Our mean Fe abundance is 7.62. Yang, Chen \& Zhao (2015) analyze high-resolution 
spectra and obtain [Fe/H] = $+0.16\pm0.06$ from four giants from a line list based on the solar spectrum. Assuming a 
solar Fe abundance of 7.50, their analysis gives the cluster abundance as 7.66.  As Yang et al. remark, 
Carrera \& Pancino obtained this same mean Fe abundance from three giants. (In the literature, the metallicity [Fe/H] 
is generally given a prominent place in text and tables. Depending on how the calibration of the Fe lines is accomplished, 
it may be necessary to add the inferred solar Fe abundance to the quoted [Fe/H] to obtain the stellar Fe abundance in order 
to effect a fair comparison between studies.)

Stellar  ages of our giants are estimated using the PARAM code with Bayesian priors
for the initial mass function (the lognormal function from Chabrier 2001)
and the star formation rate (constant). Spectroscopic effective
temperature and [Fe/H] were used as input together with the van Leeuwen's (2007)
{\it Hipparcos} parallax and the reddening-corrected $V$ magnitude. Ages are
given in Tables 1 and 2.

\section{The superclusters' origins?}

Proposals regarding a supercluster's origins  are open to test through comparisons of
 compositions and ages for supercluster members and  stars from  the putative related open
clusters. If a supercluster is dominated by stars  from an  open cluster,  there will be a  
clear uniformity for both composition and  age between cluster and supercluster.  If a 
supercluster is generated primarily by dynamical perturbations   provided to field stars by 
the Galaxy's spiral arms and central bar, the supercluster's stars will exhibit the spread in 
composition and age expected of field stars across the Galactocentric distances at which the 
perturbations are capable of directing stars to the solar neighbourhood.  Given the
presence of abundance gradients in the Galaxy,  superclusters composed of stars from different  
Galactocentric distances are expected to have a range of metallicities and, if abundance gradients 
differ  for different elements, stars in a supercluster may be expected  to have different abundance 
ratios. Additionally, the $(U,V,W)$ space occupied by a supercluster and its possibly related 
open clusters may be populated  also by stars unrelated to either stellar grouping. Some of 
these `field' stars   will be then mistakenly assigned to the supercluster. Observational errors 
in $(U,V,W)$ can serve a similar role. Our pursuit of the superclusters' origins begins with discussions 
of compositions of giants studied by us from Famaey et al.'s (2005) selections for the Hyades  Sirius superclusters.

Identification of Hyades supercluster giants with the Hyades and Prasepe open clusters presents challenges. 
The number of  giants assigned by Famaey et al. (2005) to the supercluster far exceeds the expected number 
of stars shed by the two open clusters unless their original populations were far in excess of the present 
populations. Furthermore, in our sample of giants from the Hyades supercluster, just three and possibly four 
of the giants have an age consistent with that of the Hyades open cluster.  Of the quartet, only two have the 
[Fe/H] of the Hyades giants. All other giants are considerably older than the two open clusters and about half 
have a lower metallicity than the Hyades and Praesepe cluster giants. A similar conclusion covers the selection 
of giants in the Sirius supercluster with  main sequence stars in the U Ma cluster suggesting an age of 300 Myr 
and a metallicity of [Fe/H] $\simeq 0.0$. All of our giants in the Sirius supercluster appear older than 300 Myr. 
The three youngest stars with ages of  0.7$\pm$0.2 Gyr have [Fe/H] $\simeq +0.12$ and compositions (i.e., [El/Fe]) 
similar to the rest of the sample. (There are no giants securely identified with the U Ma open cluster.)

Independent of age and  chemical composition estimates, our samples of giants cannot be traced back to a chemically 
homogeneous open cluster because of the large spread in composition for each supercluster. Giants from the Hyades and 
Sirius superclusters each span a range in [Fe/H], as shown by the histograms in Figure 2, with the former showing the 
larger range. The principal difference between the two histograms is the high [Fe/H] tail to the Hyades supercluster's 
histogram which is absent from the Sirius supercluster's histogram. Curiously, this tail is centered on the [Fe/H] of 
the giants from the Hyades and Praesepe open clusters. Although this tail might suggest the presence of Hyades 
open cluster stars among the Hyades supercluster giants, the majority of these giants are much older than the open cluster.

In short, very few of the giants in our selection of stars attributed by Famaey et al. to the Hyades and Sirius superclusters 
and analyzed here can  be identified as originating from an associated open cluster.  Thus, the giants of these superclusters 
owe their origin to either perturbations exercised on residents of the Galactic thin disc and/or contamination of the supercluster 
by unrelated field stars thanks to measurement errors and possibly too generous criteria for $(U,V,W)$. In these circumstances, it 
is of interest to compare the compositions of the giants in the two superclusters with compositions of field giants.

Several large studies of field giants are available for consideration, e.g., Luck \& Heiter (2007),  Mishenina et al. (2007, 2013) 
and Luck (2015). Inspection of plots of abundance ratios [El/Fe] versus [Fe/H] show systematic differences for some elements (El) 
among these and between other surveys, particularly for heavy elements and others where few absorption lines are available. 
In order to minimize systematic differences, we adopt abundance ratios provided by an ongoing survey of giants in open clusters 
(see Reddy, Lambert \& Giridhar 2016 and papers referenced there). This choice should minimize systematic errors because the 
analyses of the  clusters and ours of the superclusters use very similar line lists and analytical techniques. The sample of 
clusters covers the range of [Fe/H] from about $-0.2$ to $0.0$ because the clusters are primarily in the anti-centre direction. 
The sample is broken into the two subsamples [Fe/H $\geq -0.10$ with 18 clusters and [Fe/H] $\leq -0.10$ with 16 clusters.  
Our samples of supercluster giants are similarly divided into subsamples.  Mean [El/Fe] and their standard deviations are given 
in Table 11 for the two samples of open clusters, three [Fe/H] groupings from the Hyades supercluster and two [Fe/H] groupings 
from the Sirius supercluster.

Table 11 may be used to address three questions with a bearing on the origins of the superclusters. First, do the giants of the 
Hyades and Sirius superclusters  have identical compositions across their common range of [Fe/H]? Second, how do the compositions 
of the superclusters relate to compositions of the chosen sample of open clusters? Third, do the superclusters at the 
solar metallicity ([Fe/H] = 0.0) have the solar composition?

The answer to the first of the three questions is that the giants of the two superclusters define a single relation for 
each [El/Fe] to within the typical standard deviation of about $\pm0.06$. There is a hint that the El/Fe] for heavy elements 
Y to Nd that [El/Fe] for the Sirius supercluster are marginally greater than for the Hyades supercluster but this is not the 
case for Eu, the quintessential $r$-process element.

Our second question concerns the relation between the two superclusters and the sample of open clusters where the latter are 
systematically metal-poor relative to the average giant of the superclusters. Inspection of Table 11 suggests the open clusters 
are slightly underabundant in Zn, overabundant in Zr, and underabundant in La and Nd relative to the sample from the two 
superclusters but the differences are comparable to the standard deviations. Among the sample of open clusters, there is a 
spread in $s$-process heavy element abundances at a given [Fe/H]. Lambert \& Reddy (2016)  attribute the spread to different 
degrees of contamination of a cluster's natal cloud by $s$-process products from AGB stars. A similar spread may be present 
among giants from the Sirius but not the Hyades supercluster.

The third question involves the comparison between the supercluster giants at [Fe/H] = 0.0 and the solar composition. For 
almost all elements, the mean [El/Fe] at [Fe/H] = 0.0 from giants in the two superclusters are within $\pm0.10$ dex 
corresponding to differences expected from the measurement errors and for 10 of the 14 elements the difference is 
within $\pm0.05$ dex (see Table 11). One obvious exception in Table 11 is Na but the overabundance of Na in giants from 
the supercluster (and the sample of open clusters) is simply the result of the first dredge-up increasing the Na abundance 
in the stellar atmosphere (Karakas \& Lattanzio (2014). Other exceptions include Y (marginally), Ba (marginally), La and  
Nd at about 0.2 dex and Eu at about 0.15 dex. Comparisons with solar abundances necessarily incorporate possible systematic 
errors included in analyses of giant stars including non-LTE effects and the neglect of stellar spots and granulation.

The answers to the three questions suggest that the supercluster's giants are drawn from the Galactic disc rather than 
their associated open clusters with distinctive compositions, i.e.,  distinguishing sets of chemical tags. This conclusion 
affirms the interpretations reached above from the spread in [Fe/H] and the ages of the supercluster giants. Given that the samples of 
supercluster giants may be contaminated by field giants unrelated to the supercluster with its particular $(U,V,W)$, the 
detailed compositions of giant members of the superclusters seem unlikely to be powerful clues to a supercluster's origin. 
Perhaps, more information may be provided by detailed analyses of main sequence members of the Hyades and the Sirius superclusters.

\section{Other analyses of the Hyades and Sirius superclusters}

 In the following, we review published results on the compositions of stars in the two superclusters and compare these 
 results with our own as we attempt to determine contamination of the superclusters by evaporation from associated open clusters.   

\subsection{The Hyades  supercluster}

The  composition of stars comprising the Hyades supercluster and their relation to the Hyades open cluster has been 
discussed previously by De Silva et al. (2011), Pomp\'{e}ia et al. (2011) and Tabernero et al. (2012).   Samples of 
supercluster stars were chosen according to different but similar $(U,V,W)$ criteria. Assessments of the contamination 
of the Hyades supercluster by stars from the Hyades and Praesepe open clusters hinge on the chemical and age homogeneity  
of the open clusters.

Chemical homogeneity for the Hyades open cluster was established at the 0.04 dex level for a selection of elements 
from Na to Nd for F-K dwarfs (Paulson et al. 2003; De Silva et al. 2006). Remarkably, Liu et al. (2016) report that 
the cluster is `chemically inhomogeneous' at the 0.02 dex level from analyses of 19 elements from C to Ba.  Even 
the 0.04 dex limit on inhomogeneities is a minor issue relative to the spread in compositions among supercluster 
members.  Comparable  investigations of chemical homogeneity across the Praesepe open cluster have not been reported 
but the  assumption of chemical homogeneity is assumed to prevail.

Ages of stars in the supercluster may be checked against the open  clusters' age for which estimates in the literature 
include 0.7$\pm$0.1 Gyr (Salaris et al. 2004) and  625$\pm$50 Myr (Perryman et al. 1998) for the Hyades and 
0.7$\pm$ 0.1 Gyr (Salaris et al. 2004) and 590 Myr (Fossati et al. 2008) for Praesepe. If evaporation from the clusters 
is the dominant contributor to a supercluster, supercluster members'  ages will be sharply peaked at the
open cluster's age. If dynamical perturbations are the key origin, stars in a supercluster will show a spread in ages.
Famaey, Siebert \& Jorissen (2008) showed that the spread in  ages  for the Hyades (and Sirius) supercluster 
main sequence members arise primarily from dynamical perturbations acting on field disc stars. This result is supported 
by our sample: just seven of the Hyades supercluster giants in Table 1 have ages of 0.6-0.7 Gyr overlapping with the 
estimates for the open cluster giants. Of this septet, only one and, perhaps,  two have the composition of the 
Hyades and Praesepe clusters.

In an investigation  similar to ours, De Silva et al. (2011) conducted an abundance analysis of 20  giants with  
mean velocities of the Hyades cluster and with an age consistent with that of the  cluster.  Just four  of the 20 stars 
had the metallicity of the Hyades cluster. De Silva et al. concluded that the quartet ``are likely to be former members of the
Hyades open cluster" but this is possibly an overestimate because the compositions (i.e., [El/Fe]) of the quartet were 
not shown to be distinctly different from that of the other (field) giants in their sample.

Pomp\'{e}ia et al. (2011) analysed 21 main sequence stars drawn from the
Geneva-Copenhagen Survey (Nordstr\"{o}m et al. 2004) with $(U,V,W)$
velocities representative of the Hyades supercluster. Spectra
were also obtained and analysed for five certain and four possible
main sequence members of the Hyades open cluster.
Abundances were obtained for Li, Na, Mg, Fe, Zr, Ba, La, Ce, Nd and Eu.
The authors'  adoption of Li as `an efficient population tagger' was
based on the observation that the lithium abundance along the Hyades 
main sequence  is a well-defined  function of effective temperature and,  
importantly,   the Hyades Li relation  evolves with age with Li depletion 
being more rapid the lower the mass of the main sequence star. The 
Li relation for the Hyades (which is identical to that of the Praesepe) is taken from 
Cummings et al. (2017) analysis of a large sample of cluster members. 
Pomp\'{e}ia et al.'s Li abundances for the five certain and four possible 
members of the Hyades open cluster fall on the Cummings et al.'s relation (Figure 3). 
Many of the 21 stars from the supercluster have a clear Li abundance deficiency 
with respect to an open cluster star of the same effective temperature (Figure 3). 
Pomp\'{e}ia et al. from a broader consideration of the elemental abundances 
considered two of the 21 stars  to be `evaporated candidates' and indeed this 
pair fall on the open cluster's Li relation. Four other supercluster candidates 
have  a Li abundance compatible with the Li relation of the Hyades open cluster 
but were not tagged as evaporated candidates by the broader consideration of the abundances.

These broader considerations indeed show that the two evaporated candidates among the 
supercluster members share the abundances of the Hyades cluster members. Especially 
interesting in this regard is the range in heavy element abundances among Pomp\'{e}a et al.'s 
main sequence stars.  There is a clear tendency, for example,  for [La/Fe] to decline with 
increasing [Fe/H]. This pattern is repeated for other heavy  elements including for Eu, 
the quintessential $r$-process element. Such a trend is now well established from many 
studies of main sequence stars in the solar neighbourhood (see, e.g., Battistini \& Bensby (2016)). 
In Figure 4, we show Pomp\'{e}ia et al.'s differential [La/Fe] and [Nd/Fe] abundances: these 
are the abundance ratios with respect to the Hyades cluster member HD 26756.  There is a 
clear correlation between $\Delta$[La/Fe] and $\Delta$[Nd/Fe]. This correlation appears 
largely restricted to heavy elements; [Mg/Fe] may be very weakly  correlated with 
[La/Fe]  (Figure 5 top panel) and [Zr/Fe] (Figure 5 bottom panel) is less strongly 
correlated with [La/Fe] than is [Nd/Fe]. These figures confirm that the two stars 
identified by Pomp\'{e}ia et al. as evaporated from the Hyades open cluster indeed 
have the composition of cluster stars. Most of the four stars identified as `possible' 
open cluster members do not  have the same composition including [Fe/H] and heavy element 
abundances as the open cluster members. Our conclusion echoing that by Pomp\'{e}ia et al. 
is that the cluster contributed no more than two stars (the evaporated candidates)  to 
this sample of Hyades supercluster main sequence stars.

A different selection of Hyades supercluster members was made by Tabernero et al. (2012) 
who analysed 62 FGK stars with $(U,V,W)$ within 10 km s$^{-1}$ of the mean velocities of 
the Hyades supercluster (Montes et al. 2001). The sample was a mix of main sequence and 
giant stars and included  three main sequence stars and the giant $\epsilon$ Tau from the 
Hyades cluster. Abundances were reported for Na, Mg, Al, Si, Ca, Sc, Ti, V, Cr, Cu, Zn, Y, Zr, Ba, Ce  and Nd.  
The Hyades cluster's Fe abundance was reported as [Fe/H] = $+0.21$ from $\epsilon$ Tau and $+0.10$ from the three dwarf members.

Supercluster stars were identified as  evaporated stars from the Hyades cluster, firstly, 
if their [Fe/H] fell within the interval $-0.05 \leq$ [Fe/H] $\leq +0.16$ and, secondly,  
if the [El/Fe] of the various elements also matched the values of Hyades cluster members. 
Tabernero et al. concluded  primarily from the [Fe/H] determinations that 28 of their sample of 
62 (i.e., 46\%) supercluster stars had been shed by the Hyades open cluster. This is likely 
an overestimate  because  the  0.20 dex width of the [Fe/H] window seems generous in light of 
the  homogeneity of  Hyades cluster and the precision of the abundance analysis.

In light of the heavy element abundances for Pomp\'{e}ia et al.'s sample (Figure 5), we present 
Figure 6 where [Ce/Fe] and [Nd/Fe] are compared for Tabernero et al' s sample of Hyades cluster 
and supercluster stars. (Giants are offset  from the sequence defined by main sequence stars 
indicating systematic differences in abundances between dwarfs and giants.) For main sequence 
supercluster stars, there appears to be a continuous run from low [Ce/Fe] and [Nd/Fe] to 
high [Ce/Fe] and [Nd/Fe] which is similar to but weaker than the trend in Figure 5.  The three 
Hyades open cluster dwarfs fall on the relation defined by  the other stars. A concentration of 
the supercluster stars with [Fe/H] within the adopted [Fe/H] window  surround the trio of 
Hyades open cluster main sequence members but some of these stars may not have the same 
abundances as cluster members. This concentration approximates the 46\% considered by 
Tabernero et al. to have belonged to the Hyades open cluster. However, it has not been 
shown either that these stars have distinctive chemical tags enabling evaporated open cluster 
stars to be distinguished from field stars or  that their ages match the age of the Hyades 
open cluster. In fact, the mean [El/Fe] for stars with [Fe/H] within the designated window for 
cluster membership and for the three open cluster dwarfs is within a standard deviation of 
$\pm0.10$ dex of solar values (i.e., 0.0) for all elements but for V which is noted as a 
problematic element by Tabernero et al.  A comparable remark applies to the comparison between 
our [El/Fe] for the Hyades supercluster giants and Tabernero et al.'s results for their main sequence stars.

In seeking to apply the lithium abundance test employed by Pomp\'{e}ia et al., we searched for 
Li abundance determinations for the main sequence stars in Tabernero et al.'s sample. 
(Thanks to a giant's deep convective envelope causing severe Li dilution, Li is not an 
effective chemical tag among giants.)  Through {\it SIMBAD}, we searched for determinations 
of the Li abundance. Determinations or upper limits were found for 45 stars of which 25 satisfied 
and 20 failed to meet Tabernero et al.'s [Fe/H] constraint to be considered to be an 
evaporated cluster member.  Figure 7 shows the Li abundances as a function of effective 
temperature. With a single exception of V686 Per (Xing \& Xing 2012) with an  abnormally 
high Li abundance, the smooth run of Li abundance with effective temperature includes 
essentially all stars irrespective of their [Fe/H] (i.e., members with [Fe/H] within 
the chosen window and nonmembers with [Fe/H] below and above the [Fe/H] -- compare 
Figure 3 and Figure 7). This result may confirm that all the main sequence stars have a 
similar age.  However, lithium depletion is not particularly rapid. Bubar \& King (2010) 
present Li-T$_{\rm eff}$ plots for the Pleiades, Hyades, NGC 752 and M 67. At 6000 K, 
for example,  the Li abundance in the Hyades main sequence members is 2.8 at the age of 
625 Myr and has dropped to only 2.6 in NGC 752 at the age of 2.5 Gyr and to 2.4 in M 67 at 
the age of 5 Gyr. Thus, lithium is not an especially precise chronometer.

In summary, Tabernero et al.'s selection of  Hyades supercluster main sequence candidates 
may include  stars evaporated from the Hyades and Praesepe open clusters. But the adopted 
[Fe/H] window seems too relaxed to ensure that only stars from the chemically homogeneous 
Hyades (and Praesepe) clusters are securely identified in the absence of chemical tags 
distinguishing cluster from field stars. The Li abundances suffice to eliminate a couple 
of stars within the adopted [Fe/H] range  as older than the Hyades cluster. Tighter $(U,V,W)$ 
criteria should be helpful in identifying  evaporated stars within the supercluster. 

\subsection{The Sirius supercluster and the U Ma cluster}
 
The Sirius supercluster with kinematic brethren across the sky is associated  with ``a compact nucleus, 
similar in size to an ordinary loose galactic cluster" (Roman 1949)  in the U Ma constellation. 
Main sequence stars have been identified with the nucleus (i.e., the U Ma open cluster) -- see 
Roman (1949, Tables 11 and 14), Soderblom \& Mayor (1993, Table 6) and King et al. (2003,  Table 5).  
Recent abundance analyses of stars in the U Ma nucleus show a near-solar metallicity, e.g., Monier (2005) 
from three F stars obtained a mean [Fe/H] of $-0.10$, Ammler-von Eiff \& Guenther (2009) from four F stars 
obtained a mean [Fe/H] of $-0.04$ and Tabernero et al. (2017) from three F stars got a mean [Fe/H] of $-0.07$.

As an extension of these concordant abundance analyses, it is of interest to explore the possible relationship 
between the open cluster and the Sirius supercluster. King \& Schuler (2005) compiled results from the literature 
and analysed several additional stars to provide spectroscopic [Fe/H] for 17 main sequence stars. This sample 
with membership probabilities from King et al. (2003) included two from the U Ma nucleus and fifteen attributed 
to the supercluster including stars very far from the U Ma nucleus.  The mean [Fe/H] for the 17 is 
[Fe/H] $= -0.06\pm0.07$. In the $(U,V)$ plane, the sample is centered within a few km s$^{-1}$ of the   
mean $(U,V)$ of the nucleus at $(+13.9\pm0.6,+2.9\pm0.9)$  with just a couple of outliers. The Li abundances 
as a function of effective temperature followed  the expected relation for a population of stars having the 
age of the U Ma nucleus.  In addition, the  selected member stars fell on the predicted isochrone in a 
colour-magnitude diagram. Since these supercluster members  and the U Ma cluster have the same metallicity, 
an impression is given that the supercluster and open cluster are intimately related.

Ammler-von Eiff \& Guenther (2009) compiled a list of supercluster members from primarily kinematic 
assessments, particularly those by Montes et al. (2001) and King et al. (2003). Iron and magnesium 
abundances were obtained from high-resolution spectra for 17 solar-like stars, four in the nucleus 
and thirteen in the supercluster. Selected stars were  tightly clustered around the $(U,V)$ velocities 
of the nucleus, say within  $\pm2$ km s$^{-1}$ of the centroid  $(+13,+3)$. Four stars in the 
U Ma nucleus gave [Fe/H] $= -0.04\pm0.08$ and [Mg/Fe] $= -0.02\pm0.05$.  Thirteen stars in the 
Sirius supercluster gave [Fe/H] $= -0.04\pm0.04$ and [Mg/Fe] $= +0.01\pm0.03$. Thus, this study 
indicates that the U Ma nucleus and the supercluster, as so tightly defined by $(U,V)$, have 
identical  abundances of  Mg and Fe to within the tight measurement uncertainties. Also,  the 
Li abundances are as anticipated for a population of young coeval main sequence stars.  Thus, 
these stars with common Galactic velocities but with positions spread across the sky  may have 
been shed by  the U Ma nucleus.

Tabernero et al. (2017) chose main sequence stars belonging to the Sirius supercluster  using  
more relaxed kinematical criteria  than used by King \& Schuler and by Ammler-von Eiff \& Guenther, 
namely, a star was accepted as a  member if its $(U,V,W)$ were within 10 km s$^{-1}$ of the mean 
velocities of the U Ma nucleus as determined by King et al. (2003). High-resolution spectra 
provided abundances for 20 elements for  45 main sequence stars including three from the 
U Ma nucleus. Tabernero et al.'s  sample includes eight stars analysed by Ammler-von Eiff \& Guenther. 
The two analyses yield consistent results for [Fe/H] and [Mg/Fe]: mean [Fe/H] are $-0.03\pm0.06$ 
and $-0.04\pm0.05$ from Ammler-von Eiff \& Guenther and Tabernero et al., respectively. Similar 
agreement occurs for [Mg/Fe].

For Tabernero et al.'s sample with its relaxed criteria for $(U,V,W)$, metallicities [Fe/H] were 
not solely concentrated on the [Fe/H] of the U Ma cluster but ranged from $-0.37$ to 
$+0.23$ (see their Figure 2), i.e., a much broader range than found by King \& Schuler and  
Ammler-von Eiff \& Guenther. Inspection shows that [Fe/H] declines with decreasing 
effective temperature: 26 stars hotter than 5500 K give [Fe/H] $= +0.05\pm0.05$ and 
eight stars cooler than 5500 K give [Fe/H] $= -0.06\pm0.04$. By inspection of their 
Figure 6, it appears that most elements show a rather similar decrease in [El/H] with 
temperature and, thus, this systematic effect may have little influence on the ratios [El/Fe].  
The three stars from the U Ma nucleus (open cluster) and the supercluster stars with the 
[Fe/H] of the U Ma nucleus have very similar [El/Fe]: differences between the supercluster 
and the nucleus are within $\pm0.07$ dex for the great majority of the elements.  
Tabernero et al. considered stars with [Fe/H] between $-0.10$ and $+0.12$ to be  
evaporated members of the U Ma open cluster. Stars with [Fe/H] outside these limits 
were considered not to have come from the cluster. The 34 stars inside the [Fe/H] 
bounds gave a mean [Fe/H] $= 0.03\pm0.06$. Members and nonmembers are not clearly 
distinguishable by their $(U,V,W)$.

For an application of the lithium test, lithium abundances were retrieved from 
{\it SIMBAD} for 32 of the sample of 45 main sequence stars. Of the 32, all but
nine had their [Fe/H] within the range adopted by Tabernero et al. for membership 
of the U Ma cluster. The stars in the U Ma cluster's nucleus are included in the sample. 
In addition,  additional supercluster members with a Li abundance determination were 
taken from Ammler-von Eiff \& Guenther.  The Li--T$_{\rm eff}$ relation is shown in 
Figure 8  together with the Li abundances for Hyades oepn cluster stars from 
Cummings et al. (2017). Maximum Li abundances for the Sirius supercluster stars run 
on the high Li abundance side of the Hyades relation below about $T_{\rm eff} \leq 5500$ K; 
the U Ma cluster is younger than the Hyades and Praesepe open clusters.  A few stars 
with [Fe/H] consistent with U Ma cluster membership have  a Li abundance  below 
the Li--T${\rm eff}$ relation for the Sirius supercluster. These Li-poor supercluster 
stars are unlikely to have come from the U Ma cluster.  Other stars have a Li abundance 
consistent with the idea that they originated in the U Ma cluster.

In summary, compositions of the main sequence stars selected by King \& Schuler (2005) and
by Ammler-von Riff \& Guenther (2009) with their tight kinematical criteria suggest that 
the Sirius supercluster with members widely distributed across the sky suggest that many 
were evaporated from the U Ma open cluster. A similar less certain conclusion seems 
supportable from the sample of main sequence stars chosen and analysed by Tabernero et al. 
with less restrictive requirements on the $(U,V,W)$  for supercluster membership.

\section{Concluding Remarks}

The investigation in this paper concerns  the origins of stars in the Hyades and the 
Sirius superclusters and, more particularly, to what degree  these superclusters 
are populated by stars evaporated from the open clusters commonly associated with 
them, that is the Hyades and Praesepe clusters with the Hyades supercluster and the 
U Ma cluster with the Sirius supercluster?   In the attempt to answer the question, 
we have determined the chemical compositions of 34 giants in the Hyades supercluster 
and 22 giants in the Sirius supercluster and rediscussed published abundance analyses 
of main sequence and giant stars of the Hyades supercluster and the putative associated 
Hyades and the Praesepe open clusters and main sequence stars of the Sirius supercluster 
and its associated U Ma open cluster.

In principle, the chemical compositions of supercluster members may test the idea that 
stars evaporated from  associated open clusters reside in a supercluster. If the idea 
is strictly valid, stars in the open cluster and supercluster will be coeval and the 
composition of stars belonging to the cluster and supercluster will be identical. 
(Given that systematic errors may afflict  abundance analyses, the  test is most securely 
applied to stars of the same type in the cluster and the supercluster analysed in the same way.) 
The test relies on two assumptions: (i)  stars presently in  and those evaporated from the 
parent cluster have identical compositions and ages and (ii)  the open cluster has distinctive 
chemical tags enabling its stars to be  distinguished from other potential contributors or 
contaminants to the supercluster.  Relevant to (i) are  observations that open clusters are 
chemically homogeneous to about 0.04 dex for all elements other than Li. In the case of 
main sequence stars, lithium may serve as a  chemical tag, as was applied by 
Pomp\'{e}ia et al. (2011) to a sample of Hyades supercluster candidates (see their Figure 11). 
Of course, this lithium test does not apply to giant stars. (The statement about chemical 
homogeneity of a cluster requires obvious qualifiers excluding such stars as the peculiar A 
and F main sequence stars.) With respect to (ii), possible contributors are field stars from 
elsewhere in the Galactic disc. Uncertainties in stellar distances, radial velocities and 
proper motions disperse stars in the $(U,V,W)$ space and may result in accidental mixing of 
stars into the supercluster from the associated open cluster and the population of field stars.  
In a sample of giants attributed to a supercluster, the fraction of stars evaporated from the 
associated open cluster is more likely to be diluted by field stars than a sample of main sequence stars.

Judged by chemical composition and age, our sample of giants from the Hyades supercluster contains 
very few and perhaps no stars from the Hyades and Praesepe open clusters. Notably, the compositions 
of six giants from the two open clusters are at the upper end of the metallicity spread of the 
Hyades supercluster. The [Fe/H] spread among the supercluster's stars is real and not a reflection 
of measurement errors; compare the spread with the uniformity of the [Fe/H] of the Hyades and of 
the Praesepe giants. Giants from the Sirius supercluster span a smaller range in [Fe/H] than their 
counterparts in the Hyades supercluster but, nonetheless,  the spread in composition and age are 
incompatible with the proposal that our sample of Sirius supercluster giants is fed to a large 
degree by evaporation from the U Ma cluster.

In contrast, analyses of main sequence members of the superclusters may suggest the presence of 
stars evaporated from the associated open clusters. The most convincing evidence is provided for 
the Sirius supercluster and the associated U Ma open cluster. When the  selection by $(U,V,W)$ is 
as tight as insisted upon by Ammler-von Eiff \& Guenther (2009) and King \& Schuler (2005) for the 
Sirius supercluster, population of the supercluster by evaporation from the U Ma open cluster becomes 
a very likely prospect, as these authors recognized. It has yet to be shown that the Hyades supercluster 
contains appreciable numbers of stars from the Hyades and Praesepe clusters. Pomp\'{e}ia et al. (2011) 
suggested pollution of the Hyades supercluster from these open clusters was minimal. Studies by 
Tabernero et al. (2012, 2017) of main sequence stars attributable to the Hyades and Sirius superclusters  
suggest that  significant fractions of the stars in the superclusters were provided by host open clusters. 
Tabernero et al. (2012) in their abstract remark that ``46\% of our candidates are members" of the 
Hyades open cluster. For the Sirius supercluster and the U Ma open cluster, Tabernero et al. (2017)  
note that 29 out of 44 (i.e., 66\%) of their supercluster members are ``likely to originate from a 
dispersing cluster" (i.e., U Ma). These estimates of open cluster contamination of the two superclusters 
depend  on the width of the [Fe/H] ($\simeq 0.2$ dex) windows taken as consistent with the [Fe/H] of the 
open cluster  which seem  generous assessments of measurement errors and far exceed the small dispersion 
in [Fe/H] within an open cluster. These windows are also contaminated as a result of the more relaxed 
the $(U,V,W)$ windows.

To refine further our understanding of the origins of the Hyades and Sirius superclusters will call for 
tighter definitions of the $(U,V,W)$ velocities of these superclusters and their fine structure -- see 
Kushniruk et al.'s (2018) discussion based on {\it Gaia}'s astrometric data and RAVE's radial velocities 
(and other measurements). Determination of chemical compositions with a focus  on high precision should 
pay particular attention to the heavy elements (say Y to Eu) and to Li. With  trigonometrical parallaxes 
from {\it Gaia}  and stellar evolutionary tracks it may be possible to invoke age as an important parallel 
indicator  for many more main sequence stars than at present. Refinement of a supercluster's stars with 
regards to membership, kinematics, detailed composition and ages  should aid in determining the supercluster's 
origin and whether certain open clusters have contributed to the supercluster.

\section{Acknowledgments}

DLL thanks the Robert A. Welch Foundation of Houston, Texas for support through grant F-634. We thank the referee for a constructive report.

\begin{figure*} 
\centering
\includegraphics[width=16cm, height=14cm]{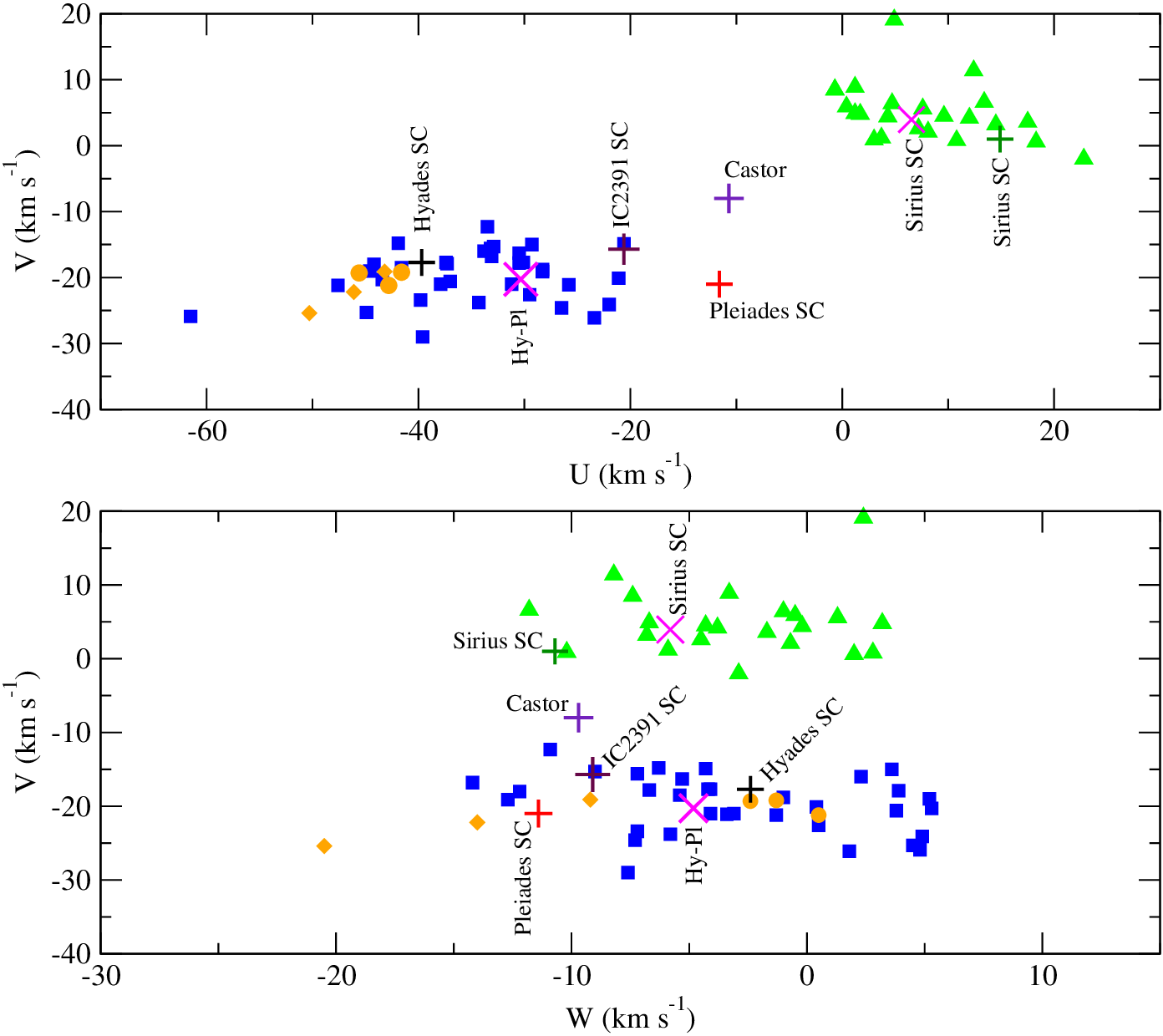}
\caption{The $(U,V)$ and $(W,V)$  plots. Blue squares and green triangles represent the observed giants from 
the Hyades and Sirius superclusters, respectively.  Orange circles and orange diamonds represent the 
Hyades open cluster giants and Praesepe open cluster giants, respectively. The + symbols represent 
the position of five superclusters or moving groups given in 
Montes et al. (2001): Sirius, Castor, Hyades, IC 2391 and the Local Association. The Magenta crosses represent the 
position of the Hyades and Sirius superclusters as given in Famaey et al. (2005). }

\end{figure*}

\clearpage

\begin{figure*} 
\centering
\includegraphics[width=16cm, height=12cm]{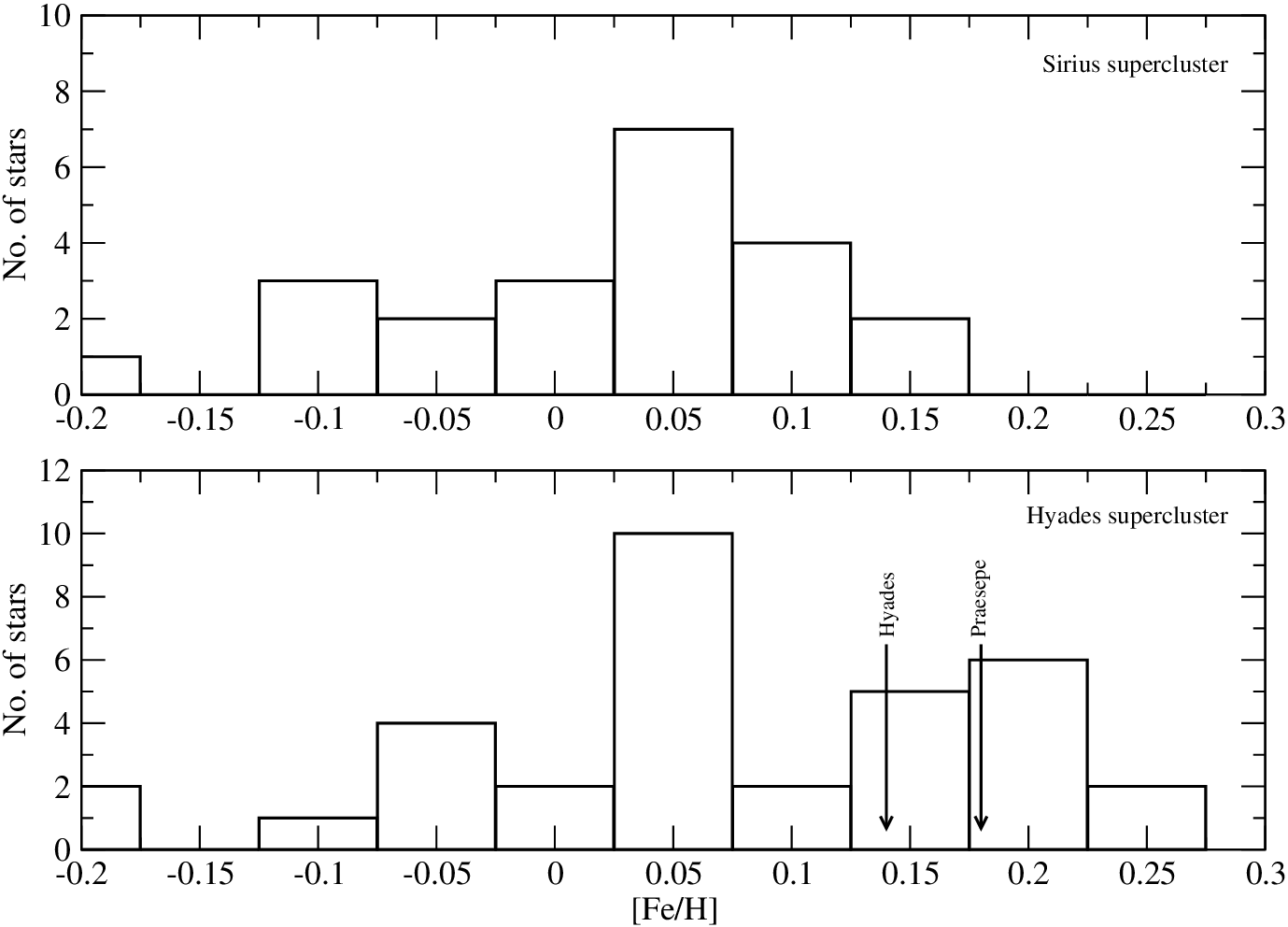}
\caption{Histograms for [Fe/H] for the sample of the 34 giants from the Hyades supercluster (bottom panel) 
and the 22 giants from the Sirius supercluster (top panel). Mean abundances of the giants from the Hyades 
and Praesepe open clusters are marked by downward pointing arrows in the bottom panel.}
\end{figure*}

\clearpage

\begin{figure*}
\centering
\includegraphics[width=16cm, height=12cm]{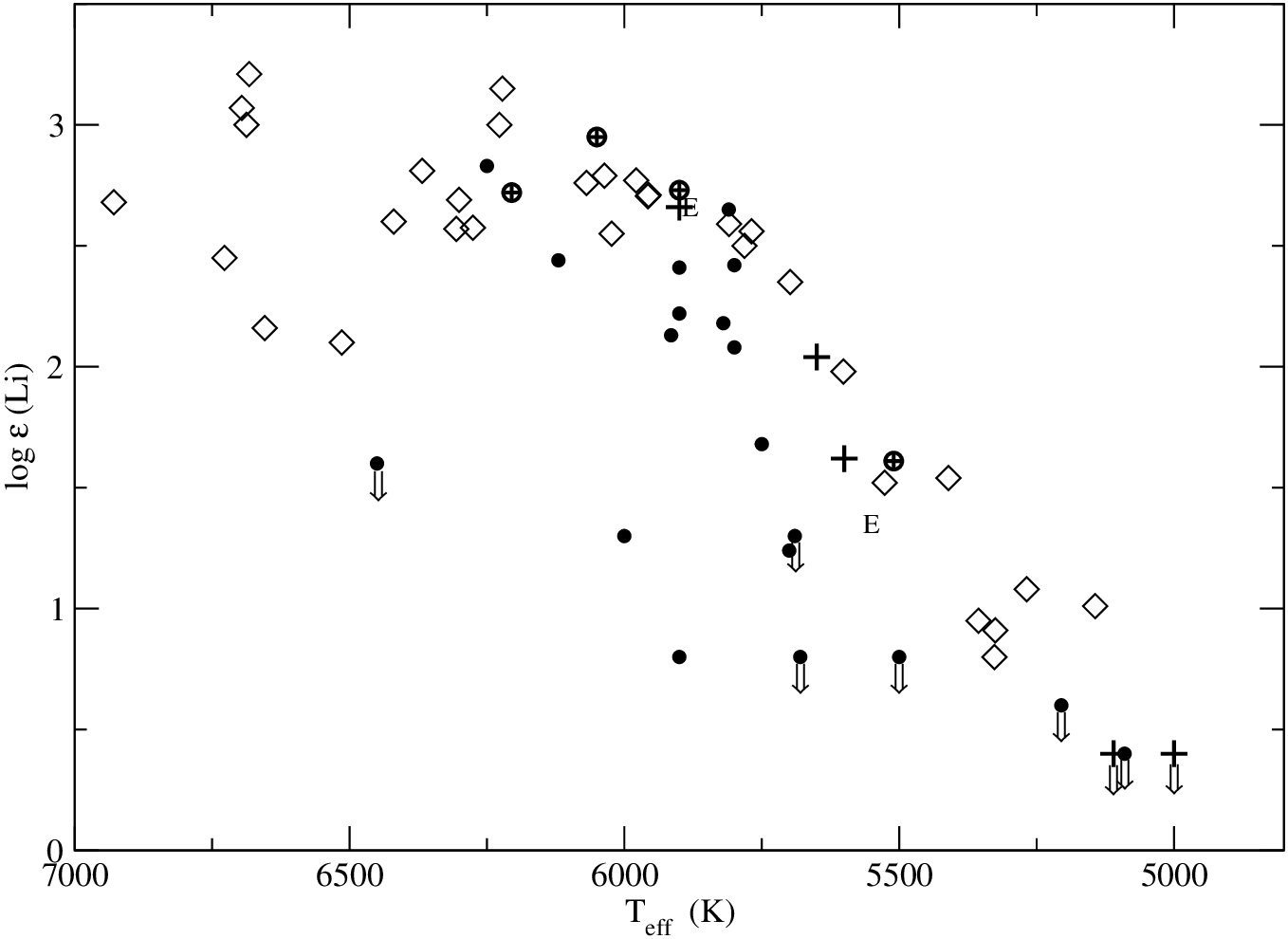}
\caption{The Lithium abundance $\log\epsilon$(Li)  and effective temperature T$_{\rm eff}$ for 
Hyades supercluster main sequence candidates selected and analysed by Pomp\'{e}ia et al. (2011). 
Lithium abundances for members of the Hyades (diamonds) open cluster are 
taken from Cummings et al. (2017). Stars designated by Pomp\'{e}ia et al. as `certain members' 
of the Hyades open cluster are represented by the + symbol, their possible members of the open cluster are 
represented  by a encircled +, and two possible evaporated cluster stars are shown by the letter E.  
Supercluster stars which by their (lower) Li abundance have not been provided by the open clusters are 
shown as filled circles. The down arrows mark upper limits on the Li abundance }
\end{figure*}

\clearpage

\begin{figure*}
\centering
\includegraphics[width=16cm, height=12cm]{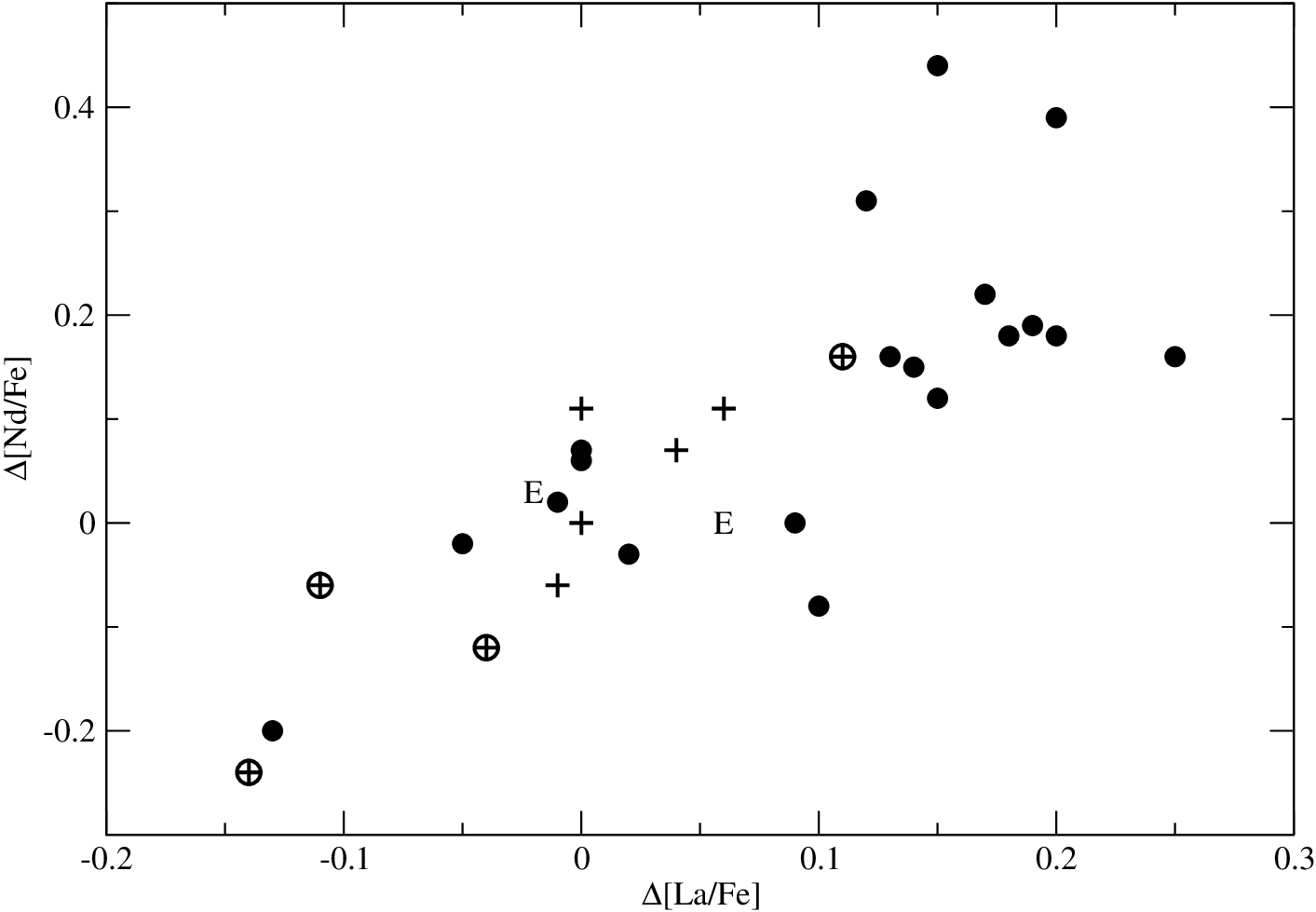}
\caption{The run of the differential abundance $\Delta$[Nd/Fe[ with $\Delta$[La/Fe]  for the stars in the 
Hyades open cluster and supercluster analysed by Pomp\'{e}ia et al. (2011). Abundance ratios are referenced 
to the values obtained for the Hyades cluster member HD 26756.  The symbols distinguish certain (+ symbol) 
and possible (encircled +) Hyades open cluster members, possible evaporated cluster stars  (capital E) among 
the supercluster's members and other members of the supercluster (filled circles). }
\end{figure*}

\clearpage

\begin{figure*}
\centering
\includegraphics[width=16cm, height=12cm]{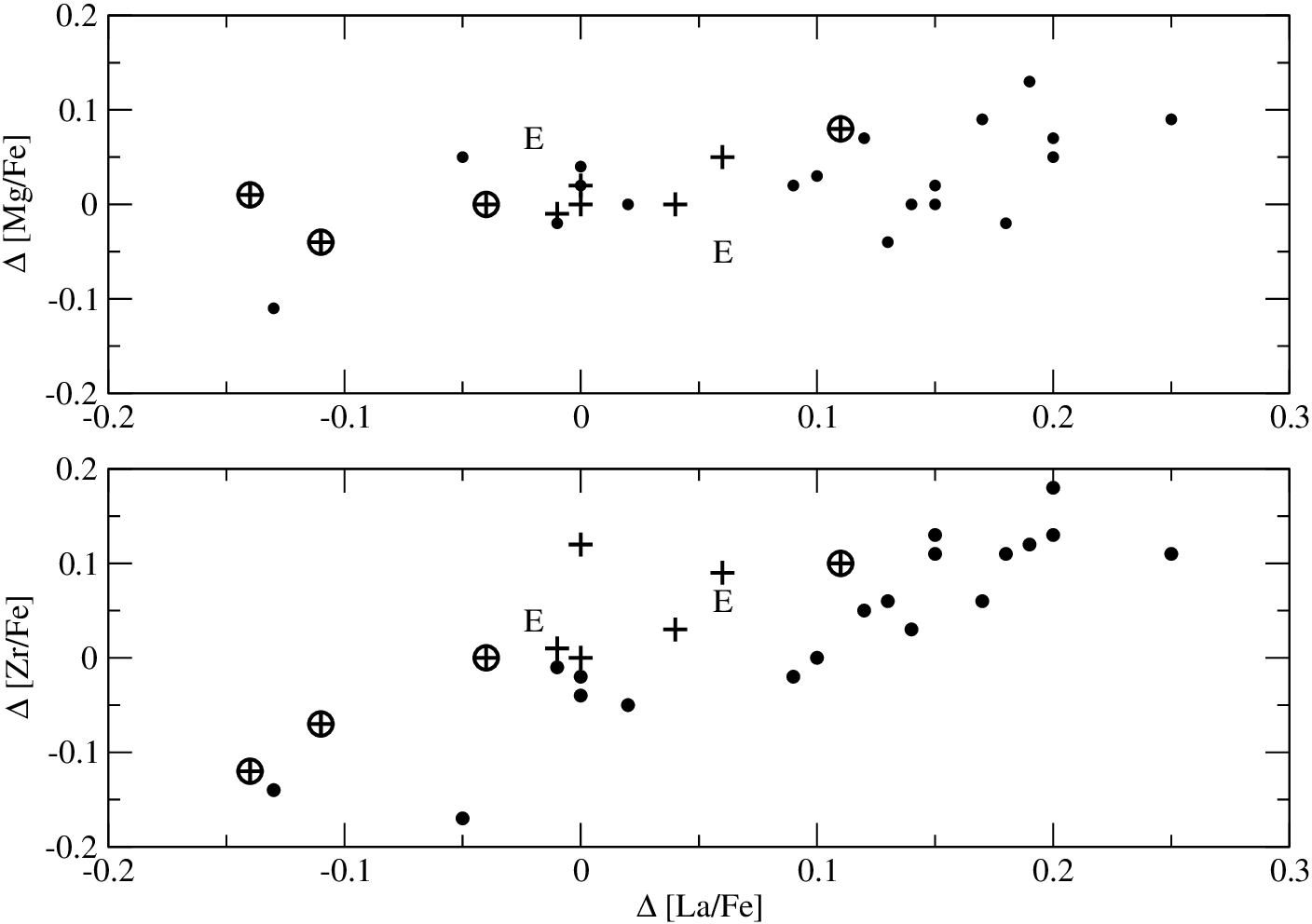}
\caption{The run of the differential abundance $\Delta$[Mg/Fe] (top panel) and $\Delta$[Zr/Fe] (bottom panel) 
with $\Delta$[La/Fe] for the stars in the Hyades open cluster and supercluster analysed by Pomp\'{e}ia et al. (2011). 
Abundance ratios are referenced to the values obtained for the Hyades cluster member HD 26756.  The symbols 
(see Figure 4) distinguish certain and possible Hyades open cluster members,  possible evaporated cluster stars 
among the supercluster's members and other members of the supercluster. }
\end{figure*}
 
\clearpage

\begin{figure*}
\centering
\includegraphics[width=16cm, height=12cm]{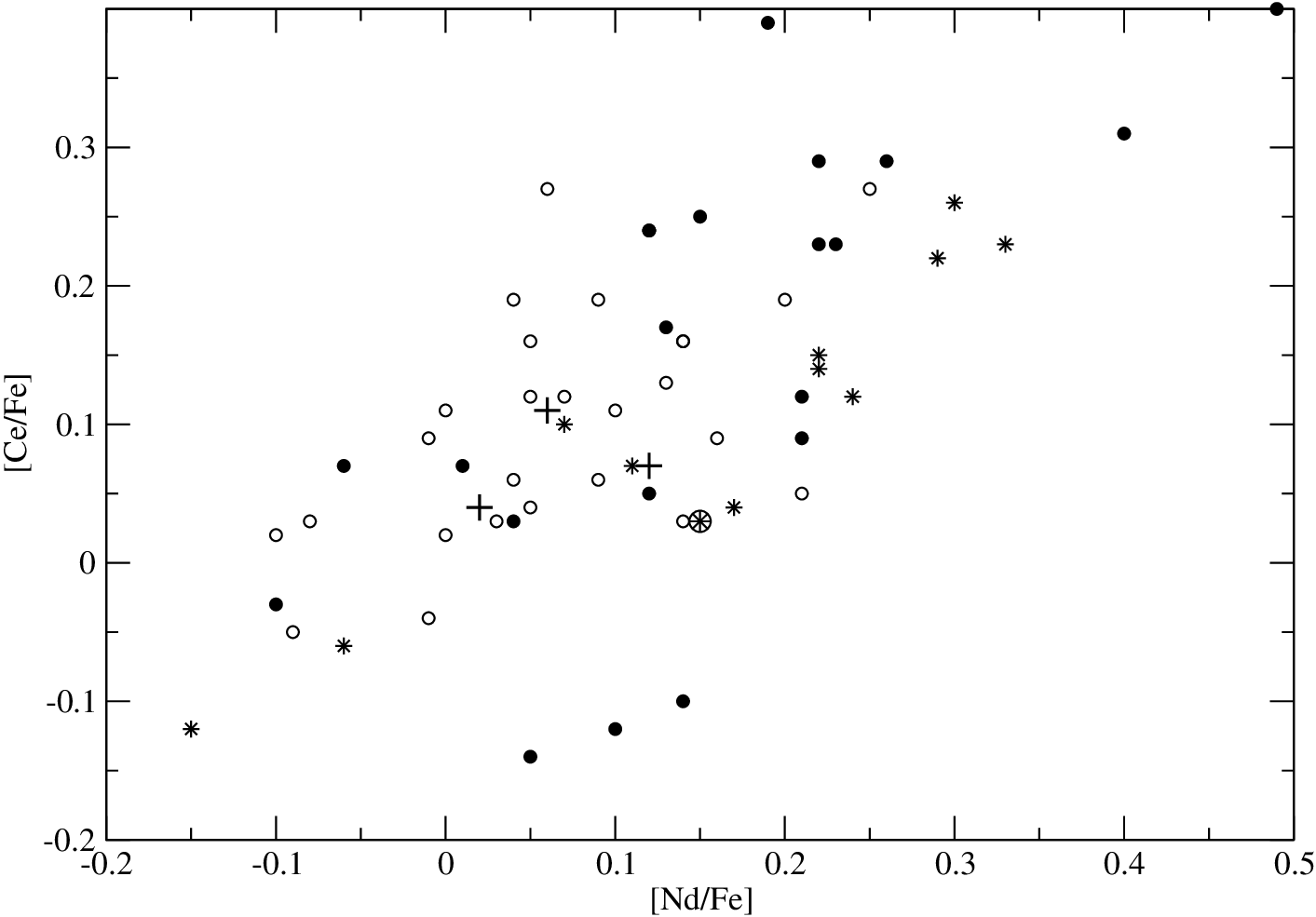}
\caption{The run of [Ce/Fe] with [Nd/Fe] for Hyades open cluster and supercluster stars analysed by 
Tabernero et al. (2012). Hyades open cluster main sequence stars are represented by + and the cluster's 
giant $\epsilon$ Tau by an encircled star symbol. Other giant stars are represented by star symbols. Supercluster 
main sequence stars satisfying the [Fe/H] limits $-0.04 \leq$ [Fe/H] $\leq +0.16$ (i.e., potential 
evaporated stars from the Hyades open cluster) are shown by open circles.  Supercluster stars with [Fe/H] 
outside the above limits are represented by filled circles. }
\end{figure*}

\clearpage

\begin{figure*}
\centering
\includegraphics[width=16cm, height=12cm]{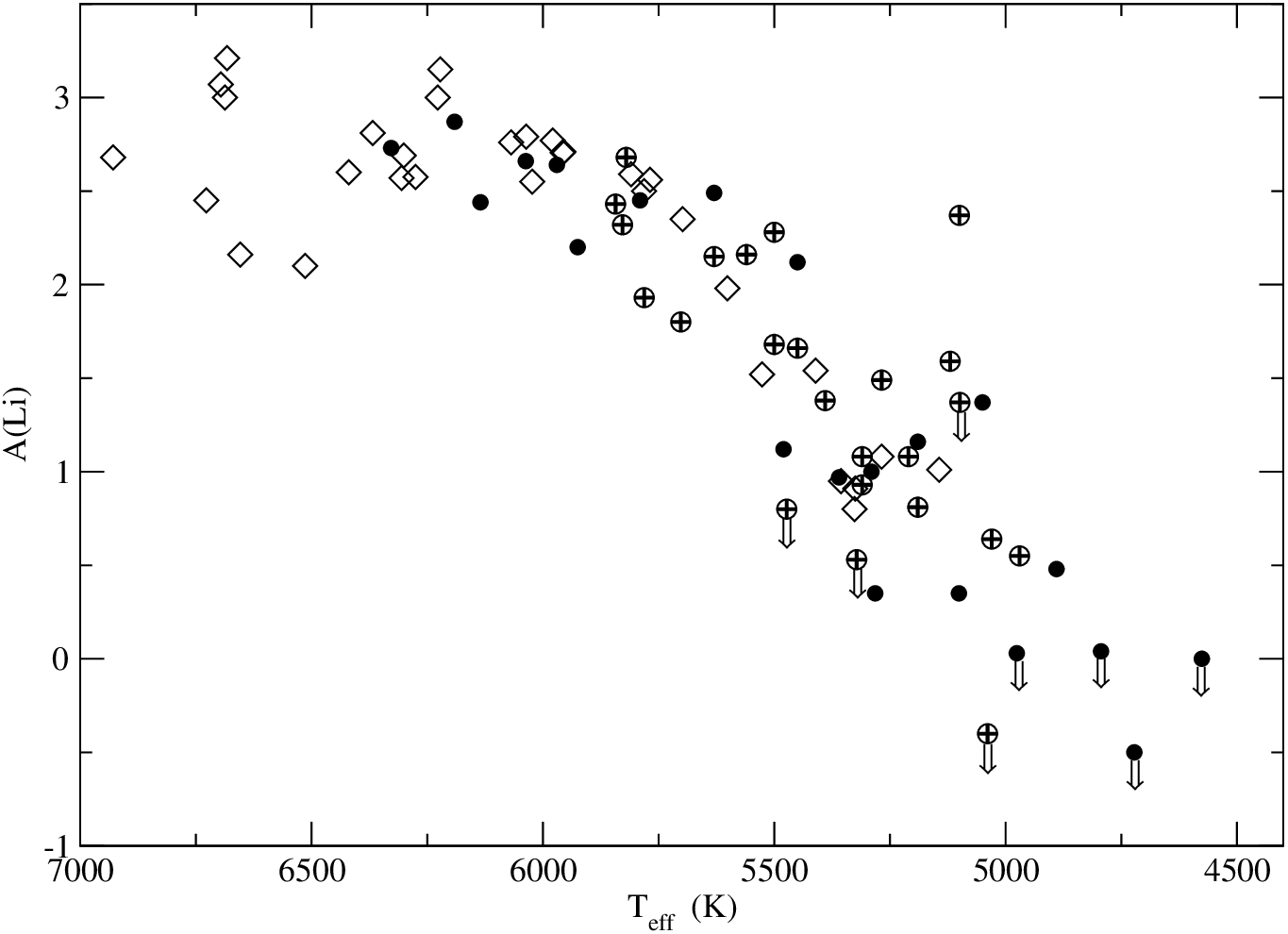}
\caption{The Lithium abundance $\log\epsilon$(Li) trend with effective temperature T$_{\rm eff}$ for 
Hyades supercluster main sequence candidates selected by Tabernero et al. (2012).  Stars satisfying their 
[Fe/H] criteria for prior membership of the Hyades open cluster are represented  by a encircled +. Stars 
failing the [Fe/H] criteria are shown as filled circles.  The diamond symbol shows the run of the Li abundance 
obtained by Cummings et al. (2017) from main sequence members of the Hyades open cluster; Hyades and Praesepe open clusters 
show essentially the same decline with effective temperature. The down arrows mark upper limits on the Li abundance }
\end{figure*}

\clearpage

\begin{figure*}
\centering
\includegraphics[width=16cm, height=12cm]{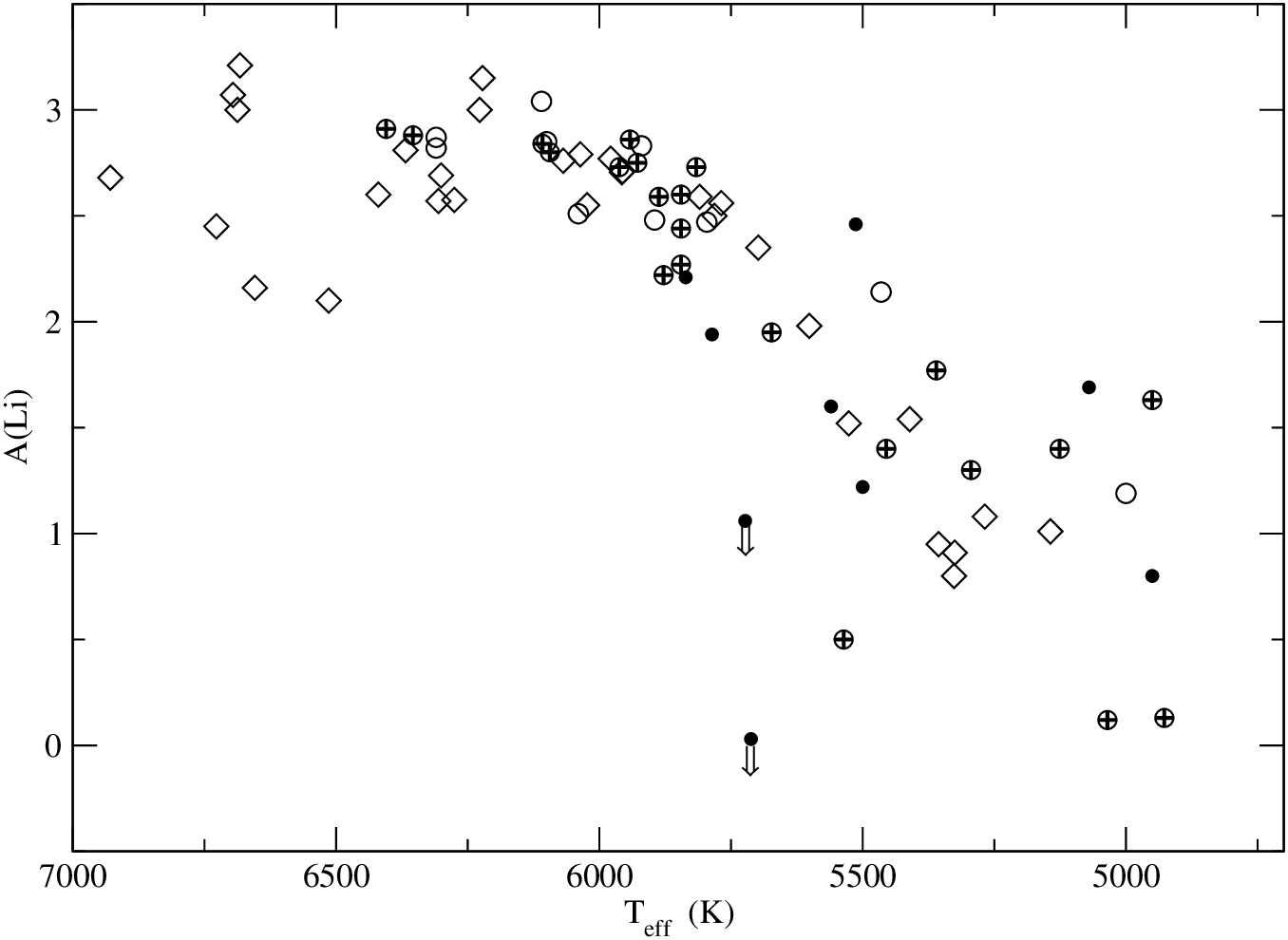}
\caption{The Lithium abundance $\log\epsilon$(Li) trend with effective temperature T$_{\rm eff}$ for 
Sirius supercluster main sequence candidates  analysed by Tabernero et al. (2017).  Stars satisfying 
their [Fe/H] criterion to have been evaporated from the U Ma open cluster are represented  by a 
encircled +. Stars failing that [Fe/H] criterion are shown as filled circles. Additional stars assigned 
to the supercluster  with a Li abundance determination by Ammler-von Eiff \& Guenther (2009) are shown by 
open circles. Abundances for Hyades open cluster main sequence stars (diamond) are taken from 
Cummings et al. (2017). The down arrows mark upper limits on the Li abundance}
\end{figure*}

\clearpage

\onecolumn

\begin{longtable}{lccccccc}
\caption{Derived Atmospheric Parameters of giants from the  Hyades supercluster.}  \\
\hline  
Star  & ($T_{\rm eff}$)$_{V-K}$ & ($T_{\rm eff}$)$_{\rm spec}$  & log $g$ (phot) & log $g$ (spec) 
                                                                           & $\xi_{t}$ & [M/H]$_{model}$ & Age \\
      & K      &    K       &    cm s$^{-2}$  &   cm s$^{-2}$   & km s$^{-1}$ &  dex  & Gyr

 \\ \hline \hline 
      
HIP1421   & 5257  & 5040  &  2.81 0.04  & 2.95     & 1.52 & +0.05       & 1.1 0.2   \\
HIP2320   & 4735  & 4870  &  2.84 0.04  & 2.78     & 1.60 & +0.25       & 1.4 0.2   \\
HIP6939   & 4651  & 4570  &  2.45 0.08  & 2.35     & 1.65 & +0.23       & 4.8 2.5   \\ 
HIP13887  & 4898  & 5020  &  2.84 0.04  & 2.80     & 1.49 & +0.10       & 1.2 0.2   \\
HIP17752  & 4689  & 4740  &  2.74 0.15  & 2.52     & 1.46 & -0.19       & 5.8 2.4    \\
\hline
\end{longtable}

\clearpage

\begin{longtable}{lccccccc}
\caption{Derived Atmospheric Parameters of giants from the Sirius supercluster.}  \\
\hline  
Star  & ($T_{\rm eff}$)$_{V-K}$ & ($T_{\rm eff}$)$_{\rm spec}$  & log $g$ (phot) & log $g$ (spec) 
                                                                           & $\xi_{t}$ & [M/H]$_{model}$ & Age \\
      & K      &    K       &    cm s$^{-2}$  &   cm s$^{-2}$   & km s$^{-1}$ &  dex  & Gyr

 \\ \hline \hline 

HIP7668  & 4276 &  4210 &  1.68 0.14  & 1.55  &  1.82   &  -0.08  & 3.0 2.6     \\
HIP26711 & 4512 &  5020 &  2.64 0.09  & 2.52  &  1.70   &  +0.14  & 0.6 0.2     \\
HIP28556 & 4563 &  4510 &  2.30 0.12  & 2.21  &  1.56   &  +0.07  & 3.0 2.3     \\
HIP35431 & 5081 &  5100 &  2.72 0.08  & 2.70  &  1.43   &  +0.01  & 0.8 0.2     \\
HIP37030 & 4410 &  4490 &  2.01 0.16  & 1.98  &  1.65   &  -0.09  & 1.1 1.0      \\
\hline
\end{longtable}

\clearpage

\begin{longtable}{lcccccccc}
\caption{The line data adopted for the current study.} \\
 \hline
Species & Wavelength &   LEP &  log $gf$  & W$_{\lambda \odot}$ &  log $\varepsilon_{\odot}$ &  log $\varepsilon \pm \sigma $  & log $\varepsilon \pm \sigma $  &  Difference  \\
        &            &       &            &                       &                            &     current study                      &  Asplund et al.(2009)         & current - Asplund  \\
        &   (\AA)   &  (eV)  &            &    (m\AA)             &      dex                   &     dex                                &   dex                          & dex     \\
        \hline

Na I    & 6154.230 & 2.100 & -1.550 & 36.6 & 6.26  &  6.27 $\pm$ 0.02   & 6.24 $\pm$ 0.04  &  +0.03   \\ 
        & 6160.750 & 2.100 & -1.250 & 56.5 & 6.28  &                    &                  &          \\
Mg I    & 5711.090 & 4.340 & -1.730 & 104.1 & 7.57 &  7.60 $\pm$ 0.03 	& 7.60 $\pm$ 0.04  & 0.00     \\ 
	& 6318.720 & 5.110 & -1.950 & 44.5 & 7.57  &                    &                  &          \\
\hline \hline

\end{longtable}

\clearpage

\begin{landscape}
\begin{longtable}{lcccccccc}
\caption{Derived abundances of giants from the  Hyades supercluster. } \\

\hline \\

Star      &  [FeI/H] $\sigma$ N &    [FeII/H] $\sigma$ N &  [Na/Fe] $\sigma$ N  &     [Mg/Fe] $\sigma$ N      &   [Al/Fe] $\sigma$ N     &   [Si/Fe]  $\sigma$ N   &   [Ca/Fe] $\sigma$ N    &   [Sc/Fe] $\sigma$ N  \\ \\  \hline

HIP1421   &  0.05 0.08 83 &  0.03 0.05 12  &  0.27 0.01 2   &  0.01 0.05 4   &  0.07 0.06 4  &  0.07 0.06 5  &  0.05 0.07 8  &  -0.02 0.04 6   \\
HIP2320   &  0.26 0.08 83 &  0.24 0.08 11  &  0.28 0.01 2   &  0.02 0.05 4   &  0.14 0.04 4  &  0.09 0.06 5  &  0.00 0.06 8  &  -0.04 0.04 6   \\
HIP6939   &  0.23 0.10 78 &  0.23 0.06 10  &  0.33 0.04 2   &  0.07 0.03 4   &  0.24 0.06 4  &  0.12 0.09 5  &  -0.09 0.07 8 &  -0.04 0.05 6    \\
HIP13887  &  0.11 0.07 86 &  0.11 0.04 11  &  0.18 0.01 2   &  -0.05 0.05 3  &  0.05 0.05 4  &  0.04 0.07 5  &  0.04 0.05 4  &  -0.10 0.04 6    \\
\hline
\end{longtable}
\end{landscape}

\clearpage

\begin{landscape}
\begin{longtable}{lccccccc}
\caption{Derived abundances of giants from the  Hyades supercluster.} \\

\hline

\\
 Star   &        [TiI/Fe] $\sigma$  N   &  [TiII/Fe] $\sigma$ N  &  [V/Fe] $\sigma$ N  &   [Cr/Fe] $\sigma$ N &  [Mn/Fe] $\sigma$ N &  [Co/Fe] $\sigma$ N  &   [Ni/Fe] $\sigma$ N    \\ \\  \hline

HIP1421     &   0.00 0.05 8  &   -0.04 0.06 4  &   -0.01 0.04 5  &   0.02 0.07 6  &   -0.06 0.03 3  &   -0.02 0.03 4  &   -0.01 0.05 10  \\
HIP2320     &   0.00 0.07 8  &   -0.09 0.05 4  &   -0.02 0.06 5  &   0.04 0.08 6  &   -0.03 0.03 3  &   0.02 0.05 4   &   0.05 0.05 11   \\
HIP6939     &   -0.09 0.09 7 &   -0.20 0.03 4  &   -0.06 0.04 5  &   0.00 0.07 6  &   -0.09 0.10 3  &   0.08 0.05 4   &   0.09 0.04 11    \\
HIP13887    &   -0.01 0.08 8 &   -0.06 0.04 4  &   -0.07 0.02 5  &   -0.04 0.05 4 &   -0.11 0.03 3  &   -0.10 0.03 4  &   -0.00 0.05 11   \\
\hline
\end{longtable}
\end{landscape}

\clearpage

\begin{landscape}
\begin{longtable}{lccccccc}
\caption{Derived abundances of giants from the  Hyades supercluster.} \\

\hline \\

Star & [Zn/Fe] $\sigma$ N & [Y/Fe] $\sigma$ N & [Zr/Fe] $\sigma$ N & [Ba/Fe] $\sigma$ N &  [La/Fe] $\sigma$ N & [Nd/Fe] $\sigma$ N  & [Eu/Fe] $\sigma$ N \\

\\ \hline

HIP1421      &   -0.07 0.09 2   &    0.13 0.06 5  &  0.05 0.06 2   &   0.08 0.01 2  &   0.17 0.06 4  &   0.25 0.03 3  &   0.15 0.00 1  \\
HIP2320      &   -0.05 0.11 2   &    0.11 0.07 5  &  -0.09 0.13 2  &   -0.02 0.04 2 &   0.07 0.12 4  &   0.11 0.03 3  &   0.12 0.00 1   \\
HIP6939      &   0.18 0.27 2    &    0.13 0.01 3  &  -0.35 0.18 2  &   -0.19 0.01 2 &   -0.12 0.05 3 &   0.00 0.08 3  &   0.07 0.00 1   \\
HIP13887     &   -0.07 0.05 2   &    0.17 0.03 5  &  0.03 0.05 2   &   0.19 0.05 2  &   0.17 0.10 4  &   0.25 0.08 3  &   0.12 0.00 1   \\
\hline
\end{longtable}
\end{landscape}

\clearpage

\begin{landscape}
\begin{longtable}{lcccccccc}
\caption{Derived abundances of giants from the  Sirius supercluster.} \\
\hline \\
Star & [FeI/H] $\sigma$ N & [FeII/H] $\sigma$ N & [Na/Fe] $\sigma$ N & [Mg/Fe] $\sigma$ N & [Al/Fe] $\sigma$ N & [Si/Fe]  $\sigma$ N & [Ca/Fe] $\sigma$ N &   [Sc/Fe] $\sigma$ N  \\ \\  \hline
HIP7668  & -0.09 0.10 75 & -0.09 0.08 11 & 0.19 0.02 2 & -0.03 0.07 4 & 0.02 0.02 4  & 0.18 0.07 4  & -0.17 0.05 8 & -0.09 0.01 5   \\
HIP26711 & 0.14 0.07 87  & 0.13 0.06 12  & 0.21 0.01 2 & -0.08 0.04 4 & 0.00 0.02 4  & 0.04 0.05 5  & 0.03 0.05 8  & -0.07 0.03 6   \\
HIP28556 & 0.06 0.08 86  & 0.06 0.07 11  & 0.10 0.03 2 & -0.03 0.04 4 & 0.04 0.03 4  & 0.09 0.06 5  & -0.06 0.05 8 & -0.04 0.04 6   \\
HIP35431 & 0.00 0.07 86  & 0.00 0.06 12  & 0.17 0.03 2 & -0.07 0.03 2 & -0.03 0.02 4 & -0.01 0.06 5 & 0.05 0.06 8  & -0.08 0.03 4   \\
\hline
\end{longtable}
\end{landscape}

\clearpage

\begin{landscape}
\begin{longtable}{lccccccc}
\caption{Derived abundances of giants from the  Sirius supercluster.} \\
\hline \\
Star   & [TiI/Fe] $\sigma$  N & [TiII/Fe] $\sigma$ N  & [V/Fe] $\sigma$ N  & [Cr/Fe] $\sigma$ N & [Mn/Fe] $\sigma$ N &  [Co/Fe] $\sigma$ N & [Ni/Fe] $\sigma$ N    \\ \\  \hline
HIP7668  & -0.19 0.06 8 & -0.10 0.02 4 & -0.22 0.07 5 & -0.06 0.09 5 & -0.29 0.07 3 & -0.01 0.04 4 & -0.04 0.06 11  \\
HIP26711 & -0.03 0.05 6 & -0.07 0.03 4 & -0.05 0.02 5 & -0.02 0.06 6 & -0.16 0.04 3 & -0.08 0.04 4 & -0.07 0.05 11  \\
HIP28556 & -0.08 0.06 5 & -0.09 0.04 4 & -0.11 0.04 5 & -0.02 0.08 6 & -0.21 0.07 3 & -0.01 0.05 4 & -0.06 0.05 11  \\
HIP35431 & -0.05 0.04 6 & -0.01 0.06 4 & -0.04 0.04 5 & -0.01 0.06 6 & -0.22 0.10 3 & -0.07 0.05 4 & -0.11 0.06 11  \\
\hline

\end{longtable}
\end{landscape}

\begin{landscape}
\begin{longtable}{lccccccc}
\caption{Derived abundances of giants from the  Sirius supercluster.} \\

\hline \\

Star & [Zn/Fe] $\sigma$ N  & [Y/Fe] $\sigma$ N  & [Zr/Fe] $\sigma$ N & [Ba/Fe] $\sigma$ N & [La/Fe] $\sigma$ N & [Nd/Fe] $\sigma$ N  & [Eu/Fe] $\sigma$ N \\

\\ \hline

HIP7668  & 0.02 0.30 2  & 0.46 0.10 4 & -0.27 0.13 2 & 0.26 0.05 2 & 0.22 0.22 4 & 0.23 0.03 2 & 0.09 0.00 1   \\
HIP26711 & -0.06 0.07 2 & 0.14 0.09 5 & 0.14 0.02 2  & 0.30 0.03 2 & 0.18 0.08 4 & 0.31 0.05 3 & 0.10 0.00 1   \\
HIP28556 & 0.06 0.10 2  & 0.34 0.18 5 & -0.05 0.01 2 & 0.17 0.02 2 & 0.21 0.14 4 & 0.44 0.19 3 & 0.12 0.00 1   \\
HIP35431 & -0.08 0.07 2 & 0.15 0.06 5 & 0.26 0.01 2  & 0.38 0.04 2 & 0.21 0.10 4 & 0.37 0.06 3 & 0.13 0.00 1   \\
\hline
\end{longtable}
\end{landscape}

\clearpage

\clearpage

\begin{landscape}
\begin{longtable}{lcccccccc}
\caption{The elemental abundances for Hyades and Praesepe open clusters.} \\
\hline \\
            &\multicolumn{4}{c} {{ Hyades (Melotte 25) }}                   & \multicolumn{4}{c} {{ Praesepe (NGC 2632) }} \\

Entity     &  $\gamma$ Tau  &  $\delta$ Tau  & $\varepsilon$ Tau&    Mean  &  HD 73598    &   HD 73665     &   HD 73710     &   Mean \\ \\
  \hline \hline

T$_{\rm eff}$ ($spec$) (K)         &  4960     & 4950     & 4890     &            & 5000           & 5020           & 4920           &    \\
log $g$ ($spec$) (cm s$^{-1}$) & 2.68      & 2.60     & 2.55     &            & 2.70           & 2.68           & 2.58           &    \\
$\xi_{t}$  (km s$^{-1}$        & 1.61      & 1.59     & 1.70     &            & 1.58           & 1.70           & 1.70           &    \\
\hline
\end{longtable}
\end{landscape}

\clearpage

\clearpage

\begin{longtable}{lccccccc}
\caption{Mean abundances for the sample of open clusters, the Hyades and Sirius supercluster giants. 
The samples are divided into two or three [Fe/H] bins as indicated in the table and N is the number of stars in a bin.} \\
 \hline

\\
Entry   &\multicolumn{2}{c} {{ Open clusters }}   & \multicolumn{3}{c} {{ Hyades supercluster }} &   \multicolumn{2}{c} {{ Sirius supercluster }}  \\
\\
\hline \hline \\

N        &   16    &    18      &     7     &    12     &      15    &      8      &   14      \\  
{}[Fe/H] &  -0.18 0.09 & -0.05 0.05 &  -0.09 0.07 & 0.04 0.02 & 0.17 0.04 &   -0.07 0.07 & 0.07 0.04
    \\
{}[Na/Fe] & 0.24 0.09 &  0.24 0.09  &  0.11 0.03 & 0.17 0.05 & 0.27 0.06  &   0.13 0.09 & 0.16 0.06
    \\
\hline
\end{longtable}

\clearpage

\end{document}